\documentclass[conference]{IEEEtran}

\usepackage{graphicx}
\usepackage{makecell}
\usepackage{listings}
\begin{document}
%
\title{Investigating Applications on the A64FX}


\author{\IEEEauthorblockN{Adrian Jackson\IEEEauthorrefmark{1}, Mich\`{e}le Weiland, Nick Brown, Andrew Turner, Mark Parsons}
\IEEEauthorblockA{EPCC, The University of Edinburgh\\
Edinburgh, United Kingdom\\
\IEEEauthorrefmark{1}Email: a.jackson@epcc.ed.ac.uk}
}

\IEEEpubid{978-1-7281-6677-3/20/\$31.00~\copyright~2020 IEEE}

\maketitle

\begin{abstract}
The A64FX processor from Fujitsu, being designed for computational simulation and machine learning applications, has the potential for unprecedented performance in HPC systems. In this paper, we evaluate the A64FX by benchmarking against a range of production HPC platforms that cover a number of processor technologies. We investigate the performance of complex scientific applications across multiple nodes, as well as single node and mini-kernel benchmarks. This paper finds that the performance of the A64FX processor across our chosen benchmarks often significantly exceeds other platforms, even without specific application optimisations for the processor instruction set or hardware. However, this is not true for all the benchmarks we have undertaken. Furthermore, the specific configuration of applications can have an impact on the runtime and performance experienced.
\end{abstract}


%
\IEEEpeerreviewmaketitle

\section{Introduction}
There is a long history of utilisation of traditional x86 architectures from processor manufacturers such as Intel and AMD for computational simulation and machine learning applications. However, we are now entering a period where there has been a significant increase in alternative processor technologies available for a wide range of tasks. Whilst Arm-based processors have been commonplace in the mobile and low power market places, a new range of Arm-based processors designs are now reaching maturity for server-class applications; foremost amongst these are Arm based processors from manufacturers such as Marvell (ThunderX2), Ampere (eMAG), Huawei (Kunpeng 920), Fujitsu (A64FX) and Amazon (Graviton, Graviton2).

Arm-based processor designs provide manufacturers and technology companies with the ability to customise processor architectures for specific workloads or requirements, and produce custom processors at volume and much more affordably that was previously possible. Leading the way in this is Fujitsu with the A64FX processor, designed in collaboration with Riken and the Japanese research community, this processor is heavily focused on a range of computational simulation and machine learning applications important to the Japanese research community. 

Having recently debuted in the Fugaku supercomputer, the current number one system on the Top500 list~\cite{top500_fugaku} and impressively entering the list at over two times the performance of the previous number one system, the A64FX has also demonstrated impressive results in efficient computing, with a Green500 rating of 16.876 GFLOPs/watts\cite{green500_fugaku}. 

However, performance is only one aspect of a system required to deliver a usable computing platform for varied workloads. Operating system, batch system, compiler, and library support are all required to provide a usable system and to ensure applications can be easily ported to such new hardware as well as efficiently exploit it.

In this paper we will evaluate a range of common computational simulation applications on a HPC system with A64FX processors connected with the TofuD network~\cite{tofud_network}. Our paper makes the following contributions to deepening the understanding of the performance of novel HPC processors, the usability of such systems, and the suitability of building a production HPC system that is based on the Fujitsu and Arm ecosystem:
\begin{enumerate}
    \item We present and evaluate the multi-node performance of scientific applications with varying performance characteristics and compare it to the established Arm and x86 ecosystems.
    \item We evaluate the ease of porting applications onto this new system, compared with equivalent systems based on other processor technologies.
    \item We discuss the causes for the performance and scalability results that have been observed, and based on this we draw conclusions with regards to the maturity of this new wave of Arm-based systems for HPC.
\end{enumerate}

\section{Related work}
The first Arm instruction set server-class processor to be widely available for typical HPC applications was the ThunderX2 processor from Marvell. The ThunderX2 processor uses the Armv8 instruction set and has been designed specifically for server workloads, although it did not exploit the new Arm SVE vectorisation instruction set. The design includes eight DDR4 memory channels to deliver measured STREAM triad memory bandwidth in excess of 240 GB/s per dual-socket node, significantly greater than the comparable Intel processors available at the time. Evaluation of scientific applications on that platform have been documented~\cite{Arm-PASC19}~\cite{10.1007/978-3-030-43229-4_17}, demonstrating comparable performance for a range of applications when compared to similar Intel processors.  This paper extends this work by adding the novel A64FX processor architecture, and expanding the range of applications, as well as the systems, benchmarked.

This paper investigates the performance of distributed memory communications (MPI), as well as scientific applications that use MPI, on an A64FX based system, and the associated libraries required for functionality and performance. It exploits the on-board high-bandwidth memory (HBM) for application performance, the first CPU-class processor to be widely available with such memory. As such, the paper presents some of the first results for user applications on both the A64FX processor and exploiting HBM.

HPC platforms have long been evaluated using a wide range of benchmarks, each targeting a different performance aspect; popular benchmark suites include~\cite{SPECMPI}~\cite{HPCChallenge}~\cite{NASBenchmark}. These include application specific benchmarks~\cite{ARCHER_Benchmarks}, and have included benchmarking application across multiple systems~\cite{ARCHER_bench_perf_report}. In this paper we follow these common benchmarking approaches to evaluate the performance of the A64FX processor against a set of other established systems using a range of different benchmarks.

The A64FX processor has been widely described~\cite{A64FX_hot_chips}~\cite{A64FX_sim}. It consists of four Core Memory Groups (CMG) connected  together with an on-chip networks (a ring bus). Each CMG has 12 cores available to the user (although a 13th core is present, provided to run the operating system and associated functionality), a coherent shared L2 cache (8MiB per CMG, 32MiB per processor) and a memory controller. HBM is directly attached to the processor, with 8 GiB attached to each CMG, providing 256GB/s of bandwidth per CMG, or approximately 1TB/s of bandwidth for the whole processor.

\section{Benchmarking methodology} 
In order to fully evaluate the performance of the A64FX, we execute a range of benchmarks and applications that rely on the performance of different aspects of the architecture, i.e. memory bandwidth, floating point performance, network performance, etc., and compare our results with other production HPC systems. Our benchmarking methodology adheres to the following principles:

\paragraph{Reproducibility} We use process and thread pinning to cores to ensure our results are not impacted or skewed by the operating system's process/thread management policies, and are reproducible. We also list the compiler versions and flags, as well as the libraries used, in Table~\ref{tab:comp}. Benchmarks are run multiple times and any performance variation outside 5\% of the average runtime is noted in the results.
\paragraph{Applications} The benchmarks and applications chosen for this investigation cover different scientific domains, programming languages, libraries and performance characteristics. They also represent widely used real-life applications. As we are primarily interested in the single node, and multi-node performance of the applications, we disabled or reduced output I/O as much as possible to ensure the I/O characteristics of the various systems considered are not dominating observed performance.
\paragraph{Multi-node benchmarks} A range of node counts are used for most benchmarks, from 1 up to 16, allowing for the assessment and evaluation of any performance bottlenecks or benefits caused by the network or the communication libraries.
\paragraph{Performance comparison} A64FX results are compared with those from a range of different HPC systems in order to assess the relative performance. The results are generally compared on a per-node basis (rather than per-core or per-process) using the same benchmark configurations.

\section{Benchmarking systems} 
The system under evaluation, a Fujitsu system containing A64FX processors, is compared against well-established HPC system architectures. Details on the systems used for this performance evaluation activity are given below, and Table \ref{tab:node} summarises the compute node specifications.

\begin{table*}[t]
\centering
\caption{Compute node specifications.}
\label{tab:node}
\begin{tabular}{|r|c|c|c|c|c|}
\hline
 & \bf{A64FX} & \bf{ARCHER} & \bf{Cirrus} & \bf{EPCC NGIO} & \bf{Fulhame} \\
\hline
\hline
Processor & \makecell{Fujitsu \\ A64FX} & \makecell{Intel Xeon \\ E5-2697 v2} & \makecell{Intel Xeon \\ E5-2695} & \makecell{Intel Xeon Platinum \\ 8260M} & \makecell{Marvell \\ ThunderX2}  \\
& {(SVE)} & {(IvyBridge)} & {(Broadwell)} & {(Cascade Lake)} & {(ARMv8)} \\
\hline
Processor clock speed & 2.2GHz & 2.7GHz & 2.1GHz & 2.4GHz & 2.2GHz\\
Cores per processor & 48 & 12 & 16 & 24 & 32  \\
Cores per node & 48 & 24 & 36 & 48 & 64  \\
Threads per core & 1 & \textbf{1} or 2 & \textbf{1} or 2 & \textbf{1} or 2 & \textbf{1}, 2, or 4 \\
Vector width & 512bit & 256bit & 256bit & 512bit & 128bit \\
\hline
Maximum node DP GFLOP/s & 3379 &  518.4 & 1209.6 & 2662.4 & 1126.4 \\
\hline
Memory per node & 32GB & 64GB & 256GB & 192GB & 256GB  \\
Memory per core & 0.66GB & 2.66GB & 7.11GB & 4GB & 4GB   \\
\hline
\end{tabular}
\end{table*}

\paragraph{A64FX} The A64FX test system has 48 compute nodes, each with a single A64FX processor, with 48 cores available to users, running at 2.2 GHz. The processor has 32GB of HBM, and nodes are connected with the TofuD network.

\paragraph{ARCHER} This Cray XC30 system has 24 cores (two Intel Xeon 2.7 GHz, 12-core E5-2697v2 processors) and 64 GB of DDR3 memory per node (128 GB on a small number of large memory nodes). Nodes are connected by the Cray Aries network.

\paragraph{Cirrus} This SGI ICE XA system has compute nodes each with two 2.1 GHz, 18-core, Intel Xeon E5-2695 (Broadwell) series processors. They have 256 GB of memory shared between the two processors. The system has a single Mellanox FDR  Infiniband (IB) fabric.

\paragraph{EPCC NGIO} This is a Fujitsu-built system where each node has two 24-core Intel Xeon Platinum  8260M processors, running at 2.4GHz, with a total of 192 GB of DDR4 Mmemory shared between the processors. The system uses Intel's OmniPath interconnect.

\paragraph{Fulhame} A HPE Apollo 70 cluster with dual-socket compute nodes connected with Mellanox EDR IB using a non-blocking fat tree topology. Each compute node consists of two 32-core Marvell ThunderX2 processors running at 2.2GHz, and 256GB DDR4 memory. 

\begin{table*}[!htbp]
\caption{Compilers, Compiler Flags and Libraries.}\label{tab:comp}
\begin{tabular}{|r|l|l|l|}
\hline
 & \bf{Compiler} & \bf{Compiler flags} & \bf{Libraries} \\
\hline
\hline
\bf{HPCG} &  &  &  \\
\hline
A64FX & Fujitsu 1.2.24  & \makecell[l]{-Nnoclang -O3 -Kfast} & \makecell[l]{Fujitsu MPI} \\
\hline
ARCHER  & Intel 17 & \makecell[l]{-O3} & \makecell[l]{Cray MPI} \\
\hline
Cirrus & Intel 17 & \makecell[l]{-O3 -cxx=icpc -qopt-zmm-usage=high} & \makecell[l]{HPE MPI MPI} \\
\hline
EPCC NGIO & Intel 19  & \makecell[l]{-O3 -cxx=icpc -xCore-AVX512 -qopt-zmm-usage=high} & \makecell[l]{Intel MPI} \\
\hline
Fulhame & GCC 8.2 & \makecell[l]{-O3 -ffast-math -funroll-loops -std=c++11 -ffp-contract=fast -mcpu=native} & \makecell[l]{OpenMPI} \\
\hline
\hline
\bf{minikab} &  &  &  \\
\hline
A64FX & Fujitsu 1.2.25 & \makecell[l]{-O3 -Kopenmp -Kfast -KA64FX -KSVE -KARMV8\_3\_A \\
            -Kassume=noshortloop -Kassume=memory\_bandwidth \\
            -Kassume=notime\_saving\_compilation} & \makecell[l]{Fujitsu MPI \\} \\
\hline
EPCC NGIO & Intel 19 & \makecell[l]{ -O3 -warn all} & \makecell[l]{Intel MPI library} \\
\hline
Fulhame & Arm Clang 20 & \makecell[l]{-O3 -armpl -mcpu=native -fopenmp} & \makecell[l]{OpenMPI \\ ArmPL} \\
\hline
\hline
\bf{nekbone} &  &  &  \\
\hline
A64FX & Fujitsu 1.2.24  & \makecell[l]{-CcdRR8 -Cpp -Fixed -O3 -Kfast -KA64FX -KSVE -KARMV8\_3\_A \\
            -Kassume=noshortloop -Kassume=memory\_bandwidth \\
            -Kassume=notime\_saving\_compilation} & \makecell[l]{Fujitsu MPI } \\
\hline
ARCHER & GCC 6.3 & \makecell[l]{-fdefault-real-8 -O3} & \makecell[l]{Cray MPICH2 library 7.5.5}
\\
\hline
EPCC NGIO & Intel 19.03  & \makecell[l]{-fdefault-real-8 -O3} & \makecell[l]{Intel MPI 19.3 } \\
\hline
Fulhame & GNU 8.2 & \makecell[l]{-fdefault-real-8 -O3} & \makecell[l]{OpenMPI 4.0.2}
\\
\hline
\hline
\bf{CASTEP} &  &  &  \\
\hline
A64FX & Fujitsu 1.2.24 & \makecell[l]{-O3} & \makecell[l]{Fujitsu MPI \\ Fujitsu SSL2 \\ FFTW 3.3.3} \\
\hline
ARCHER  & GCC 6.2 & \makecell[l]{-fconvert=big-endian -fno-realloc-lhs -fopenmp -fPIC \\ -O3 -funroll-loops -ftree-loop-distribution -g -fbacktrace} & \makecell[l]{Cray MPICH2 library 7.5.5 \\ Intel MKL 17.0.0.098 \\ FFTW 3.3.4.11} \\
\hline
Cirrus & Intel 17 & \makecell[l]{-O3 -debug minimal -traceback -xHost} & \makecell[l]{SGI MPT 2.16 \\ Intel MKL 17. \\ FFTW 3.3.5} \\
\hline
EPCC NGIO & Intel 17 & \makecell[l]{-O3 -debug minimal -traceback -xHost} & \makecell[l]{Intel MPI library 17.4 \\ Intel MKL 17.4 \\ FFTW 3.3.3} \\
\hline
Fulhame & GCC 8.2 & \makecell[l]{-fconvert=big-endian -fno-realloc-lhs -fopenmp -fPIC \\ -O3 -funroll-loops -ftree-loop-distribution -g -fbacktrace} & \makecell[l]{HPE MPT MPI library (v2.20) \\ ARM Performance Libraries 19.0.0 \\ FFTW 3.3.8} \\
\hline
\hline
\bf{COSA} &  &  &  \\
\hline
A64FX & Fujitsu 1.2.24 & \makecell[l]{-X9 -Fwide -Cfpp -Cpp -m64 -Ad -O3 -Kfast \\ -KA64FX -KSVE -KARMV8\_3\_A \\ -Kassume=noshortloop  -Kassume=memory\_bandwidth \\ -Kassume=notime\_saving\_compilation} & \makecell[l]{Fujitsu MPI \\ Fujitsu SSL2 \\ FFTW 3.3.3} \\
\hline
ARCHER & GNU 7.2 & \makecell[l]{-g -fdefault-double-8 -fdefault-real-8 -fcray-pointer \\ -ftree-vectorize -O3  -ffixed-line-length-132} & \makecell[l]{Cray MPI library (v7.5.5) \\ Cray LibSci (v16.11.1)} \\
\hline
Cirrus & GNU 8.2 & 
\makecell[l]{-g -fdefault-double-8 -fdefault-real-8 -fcray-pointer \\ -ftree-vectorize -O3  -ffixed-line-length-132} & \makecell[l]{SGI MPT 2.16 \\ Intel MKL 17.0.2.174} \\
\hline
EPCC NGIO & Intel 18 & 
\makecell[l]{-g -fdefault-double-8 -fdefault-real-8 -fcray-pointer \\ -ftree-vectorize -O3  -ffixed-line-length-132} & \makecell[l]{Intel MPI \\ Intel MKL 18} \\
\hline
Fulhame & GNU 8.2 & \makecell[l]{-g -fdefault-double-8 -fdefault-real-8 -fcray-pointer \\ -ftree-vectorize -O3  -ffixed-line-length-132} & \makecell[l]{HPE MPT MPI library (v2.20) \\ ARM Performance Libraries (v19.0.0)} \\
\hline
\hline
\bf{OpenSBLI} &  &  &  \\
\hline
ARCHER & Cray Compiler v8.5.8 & \makecell[l]{-O3 -hgnu} & \makecell[l]{Cray MPICH2 (v7.5.2) \\ HDF5 (v1.10.0.1)} \\
\hline
Cirrus & Intel 17.0.2.174 & \makecell[l]{-O3 -ipo -restrict -fno-alias } & \makecell[l]{SGI MPT 2.16 \\ HDF5 1.10.1} \\
\hline
EPCC NGIO & Intel 17.4 & \makecell[l]{-O3 -ipo -restrict -fno-alias } & \makecell[l]{Intel MPI 17.4 \\ HDF5 1.10.1} \\
\hline
Fulhame & Arm Clang 19.0.0 & \makecell[l]{-O3 -std=c99 -fPIC -Wall} & \makecell[l]{OpenMPI 4.0.0 \\ HDF5 1.10.4} \\
\hline
\end{tabular}
\end{table*}

\section{Benchmarks}
To evaluate the overall performance of the A64FX system and place it in context with the other systems considered in this paper we ran the HPCG benchmark across the systems and evaluated the comparative performance.

\subsection{HPCG}
HPCG~\cite{Dongarra15HPCG} (High Performance Conjugate Gradients) is a HPC system benchmark designed to be more representative than the traditional High Performance LINPACK (HPL\cite{HPL}) as it has a more realistic resource usage pattern, closer to full scale HPC applications. As such, HPCG performance is influenced by memory bandwidth, floating point capability and to some extent network performance.

For the benchmarks presented here we compiled HPCG in MPI only mode with as many MPI processes used as there are cores on the node. To fit into the memory on a single A64FX node we used the following parameters for benchmark across the systems: \texttt{--nx=80 --ny=80  --nz=80 --t=3600}.

For the Fulhame and EPCC NGIO systems two different versions of the benchmark were run, the first version ({\it unoptimised}) is the standard HPCG source code, the other ({\it optimised}) were modified versions of HPCG optimised by Intel and Arm respectively for the target architectures used (although still compiled with the same compiler flags). Table~\ref{tab:hpcg_results_single_node} shows the HPCG performance for a single node across the test systems.

\begin{table}[t]
\centering
\caption{Single node HPCG performance.}\label{tab:hpcg_results_single_node}
\begin{tabular}{|c|r|r|}
\hline
\bf{System } & \bf{Performance } & \bf{\% of Theoretical}  \\
\bf{}  & \bf{(GFLOP/s)} & \bf{Peak Performance} \\
\hline
\hline
A64FX & 38.26 & 1.1 \\
ARCHER & 15.65 & 3.0 \\
Cirrus & 17.27  & 1.4 \\
EPCC NGIO (unoptimised) & 26.16 & 1.4 \\
EPCC NGIO (optimised) & 37.61 &  2.0 \\
Fulhame (unoptimised) & 23.58 & 2.0\\
Fulhame (optimised) & 33.80  & 3.0 \\
\hline
\end{tabular}
\end{table}

The table demonstrates that the A64FX processor achieves significantly higher performance (approx. $30\%$) than the unoptimised HPCG source code running on the dual-socket Cascade Lake node, whilst having the same number of cores. It also shows that the A64FX has higher performance (approx. $10\%$) than the ThunderX2 node (Fulhame) whilst having fewer cores, demonstrating the performance benefits the wider vector units and/or high bandwidth memory provided by the processors.

Both Intel and Arm have optimised versions of the HPCG source code that have been altered to use specific performance libraries or optimised versions of some of the computational kernels in the benchmark. We also ran these optimised versions and provide the performance for those versions in the table. Note, both the unoptimised and optimised versions of the benchmark on EPCC NGIO and Fulhame utilise the relevant optimised mathematical libraries (MKL on NGIO and armpl on Fulhame), so the performance difference exhibited by the optimised version of the benchmark are primarily from the changes that have been made to the computational routines.

The comparison between the optimised and unoptimised versions of HPCG on NGIO and Fulhame demonstrate the scope for performance improvements available. Given that we ran an unoptimised version of HPCG on the A64FX system there is likely to be significant scope for increasing the performance through targeted code modifications on the A64FX processor, with our comparative benchmarks suggesting $30\%$ performance improvements could be possible.

We also ran the same benchmark across multiple nodes to evaluate performance once the network is required for the calculations. Table~\ref{tab:hpcg_results_multi_node} shows the performance scaling up to 8 nodes using the same configuration for the benchmark as for the single node case. 

\begin{table}[h]
\centering
\caption{Multiple node HPCG performance (GFLOP/s).}\label{tab:hpcg_results_multi_node}
\begin{tabular}{|c|r|r|r|r|}
\hline
\bf{System} & \bf{1 node} & \bf{2 nodes} & \bf{4 nodes} & \bf {8 nodes} \\
\hline
\hline
A64FX & 38.26 &  78.94 & 157.46 & 313.50 \\
ARCHER & 15.65 & 26.25 & 55.63 & 110.52 \\
Cirrus & 17.27 & 34.26  & 68.44 & 136.06 \\
EPCC NGIO (optimised) & 37.61 & 73.90 & 147.94 & 292.60 \\
Fulhame (optimised) & 33.80 & 67.68 & 133.29 & 261.32 \\
\hline
\end{tabular}
\end{table}

We can see from the multi-node benchmarks that the A64FX nodes are still providing higher performance than the rest of the systems, with the difference between A64FX and EPCC NGIO more pronounced on multiple nodes. This demonstrates that there is no significant overhead from the network hardware or libraries on the A64FX as compared to the other systems, and indeed the network may be outperforming the EPCC NGIO system (although further in-depth analysis would be need to verify this assertion).

\section{Mini-apps}

To enable in-depth analysis of performance of the A64FX without requiring time consuming full application runs we investigated a number of {\it mini-apps}. These are benchmarking programs designed to provide representative functionality of the core components of a larger scale applications. In this paper we used two mini-apps, minikab, a parallel conjugate gradient (CG) solver, and nekbone, a representative Navier-Stokes solver (NS). The section will describe those mini-apps and the performance experienced across systems for these benchmarks.

\subsection{minikab}
The Mini Krylov ASiMoV Benchmark (minikab) program is a simple parallel CG solver developed at EPCC to allow testing of a range of parallel implementation techniques. It is written in Fortran 2008 and parallelised using MPI, as well as MPI with OpenMP. It supports a range of command-line options to test the different methods that can be used when implementing a solver:

\begin{itemize}
\item the type of decomposition;
\item the solver algorithm;
\item the communication approach;
\item the serial sparse-matrix routine in plain Fortran or implemented via a numerical library (such as MKL). 
\end{itemize}

We tested minikab using a sparse matrix (called Benchmark1) that has 9,573,984 degrees of freedom and 696,096,138 non-zero elements - the matrix represents a large structural problem.



To establish a baseline for further performance analysis, the test case was run on a single core on EPCC NGIO and Fulhame, in addition to the A64FX. As table~\ref{tab:minikab_single_core} presents, on a single core, the A64FX shows the best performance by far: it is 7\% faster than even a top of the range Intel Xeon core, and just over 2x faster than the ThunderX2.

\begin{table}[h]
\begin{center}
\caption{Single code minikab performance (runtime in seconds)}\label{tab:minikab_single_core}
 \begin{tabular}{|r|c|} 
 \hline
 CPU & Runtime (s) \\ 
 \hline\hline
 A64FX & 1182 \\
 EPCC NGIO  & 1269 \\
 Fulhame  & 2415\\
 \hline
\end{tabular}
\end{center}
\end{table}

\begin{figure}
\centering
\includegraphics[width=\linewidth]{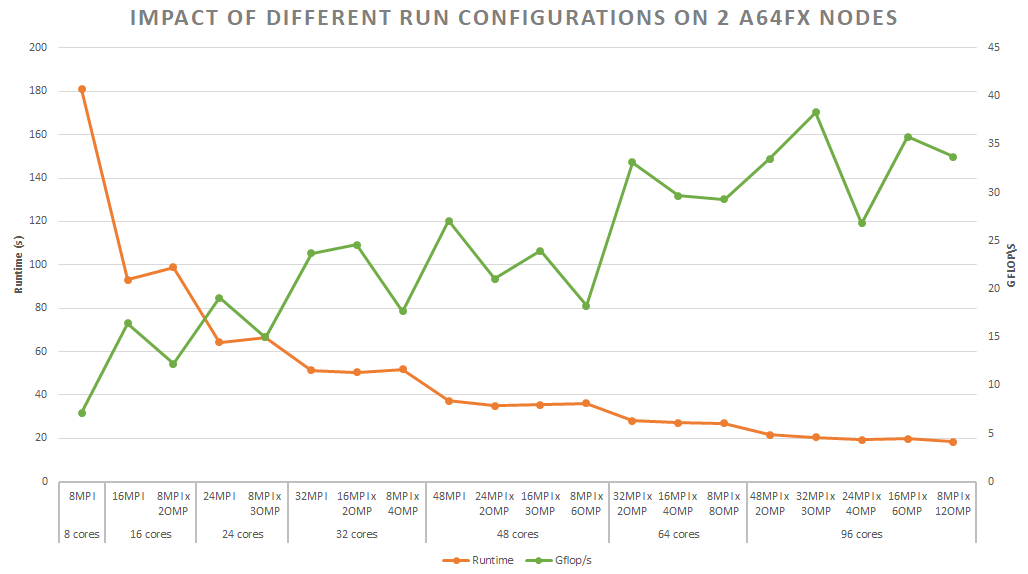}
\caption{Comparing the solver runtimes and GFLOP/s for different execution setups (plain MPI and mixed-mode MPI with OpenMP) on 2 A64FX nodes for increasing core counts. Note that the GFLOP/s are as reported by the Fujitsu profiler for the entire execution, and therefore include the setup phase, whereas the runtimes are for the solver only. In particular for higher MPI process counts, GFLOP/s may be high even though the runtime is also high.}
\label{fig:timevsflops}
\end{figure}

We also investigate the impact of different process-thread configurations for this benchmark. Our experiments, presented in Figure~\ref{fig:timevsflops}, confirm that using 1 process per CMG with 12 OpenMP threads per process gives the best performance for minikab. We compare a wide range of run configurations for 2 nodes for increasing numbers of cores used. The largest plain MPI configuration able to fit into the available memory is 48 MPI processes, i.e. under-populating the nodes by half. Unsurprisingly, the best performance is achieved when using all the cores, and out of the five options tested, using 8 MPI processes, each with 12 OpenMP threads, is fastest.

\begin{figure}
\centering
\includegraphics[width=\linewidth]{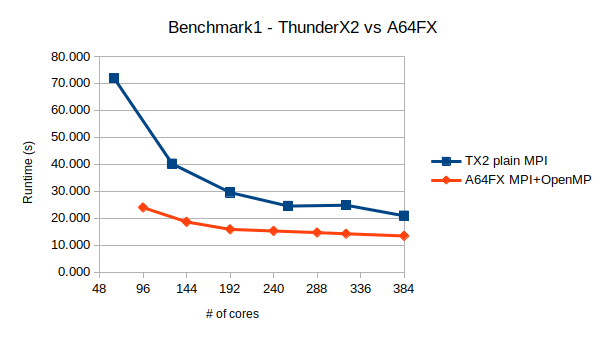}
\caption{Performance of minikab with Benchmark1 on ThunderX2 (Fulhame) and A64FX. Using up to 6 nodes on Fulhame and 8 on A64FX (strong scaling).}
\label{fig:minikab_scaling_comp}
\end{figure}

Figure~\ref{fig:minikab_scaling_comp} compares the scaling behaviour of the default setup of minikab on A64FX and Fulhame. On Fulhame, using plain MPI gives the best performance, and as memory limitations are not a concern on that system, we populated the nodes fully with MPI processes. As shown earlier, memory constraints on the A64FX mean that it is not possible to use fully populated nodes with a plain MPI configuration here. We therefore use the best performing setup in both cases as a comparison. The Fulhame results are for 1 to 6 nodes (64-384 cores) and the A64FX results are for 2 to 8 nodes (96-384 cores); three of the datapoints (192, 320 and 384 cores) match between the two systems.

We can see that the A64FX system outperforms Fulhame across the range of core counts, albeits with different numbers of nodes (i.e. on Fulhame 192 cores is 3 nodes, but 4 nodes on the A64FX). Even comparing node to node performance the A64FX is still significantly faster, although it does not scale as well as the Fulhame system.



\subsection{Nekbone}
The Nekbone mini-app benchmark captures the basic structure the Nek5000 application, which is a high order, incompressible NS solver based on the spectral element method. Nekbone solves a standard Poisson equation using a conjugate gradient iterative method with a simple preconditioner on a block or linear geometry. As a mini-app Nekbone represents the principal computational kernel of Nek5000, to enable exploration of the essential elements of the algorithmic features that are pertinent to Nek5000. 

The solution phase consists of conjugate gradient iterations that call the main computational kernel, which
accounts for over 75\% of the runtime. This kernel, \emph{ax}, performs a matrix vector multiplication operation in an element-by-element fashion. Overall each iteration of the solver involves vector operations, matrix-matrix multiply operations, nearest-neighbour communication, and MPI Allreduce operations. The linear algebra operations are performed on an element by element basis, with each element consisting of a specific polynomial order configuration (for the tests executed here we use 16 by 16 by 16). This represents a challenging computational pattern, as relatively small vector and matrix-matrix multiply operations are performed on each element, rather than a single much larger operation which libraries such as BLAS are often optimised for. Furthermore, different aspects of the kernel are bound by different limits, for instance some parts of the kernel are memory bound, whereas others are compute bound. This therefore makes it a very interesting study, exploring not only the benefits that the floating point performance of the A64FX can provide, but also whether the higher memory bandwidth can deliver benefit too.

The benchmarks are undertaken using a weak scaling methodology and leverage the largest test-case in the Nekbone repository. This corresponds to a system comprising of 200 local elements, each 16 by 16 by 16 polynomial order. Unless otherwise stated, all compilation is performed at O3, with additional architectural specific flags for optimal performance. All results reported are averaged over three runs.

\subsubsection{Node performance}

Table \ref{fig:nekbonenode} illustrates the performance in GFLOP/s of the Nekbone weak scaling experiment run across nodes of different machines which represent numerous architectures. Furthermore, we saw a significant performance improvement on the A64FX by compiling with  \emph{-Kfast}, and are denoting this as \emph{fast math} in the table, similar to \emph{-ffast-math} with GCC. It can be seen that using  the \emph{-Kfast} flag very significantly improves the performance on the A64FX, but similar compiler flags do not significantly improve performance on other architectures.

\begin{table}
\begin{center}
 \caption{Node performance of Nekbone across numerous systems}
 \begin{tabular}{|r|r|r|r|r|r|} 
 \hline
 System & \makecell[c]{Cores \\ used} & GFLOP/s & \makecell[c]{Ratio \\to A64FX} & \makecell[c]{GFLOP/s \\fast math} & \makecell[c]{Ratio \\ to A64FX} \\ 
 \hline\hline
 A64FX & 48 & 175.74 & 1.00 & 312.34 & 1.00\\
 EPCC NGIO & 48 & 127.19 & 0.72 & 90.37 & 0.29\\
 Fulhame & 64 & 121.63 & 0.69 & 132.65 & 0.42\\
 ARCHER & 24 & 66.55 & 0.40 & 68.22 & 0.21\\
 \hline
\end{tabular}
\label{fig:nekbonenode}
\end{center}
\end{table}


The table demonstrates the A64FX is outperforming all other technologies, but crucially it is the improvement memory bandwidth of this chip that makes the difference here. Furthermore, with fast maths enabled the A64FX is likely able to keep the FPUs busy with data, whilst the other technologies are not and hence likely stalling on memory access. For comparison, with a similar sized number of elements, Nekbone performance experiments explored in \cite{karp2020optimization} demonstrate approximately 200 GFLOP/s on a P100 GPU, and 300 GFLOP/s on a V100 GPU. Therefore at 312 GFLOP/s Nekbone on the A64FX with fast maths enabled is competitive against runs on a GPU, significantly outperforming a P100 and marginally faster than a V100. The \emph{-Kfast} flag is critical here, without it the performance delivered by the A64FX falls short of both GPU technologies.





\begin{figure}
\centering
\includegraphics[width=\linewidth]{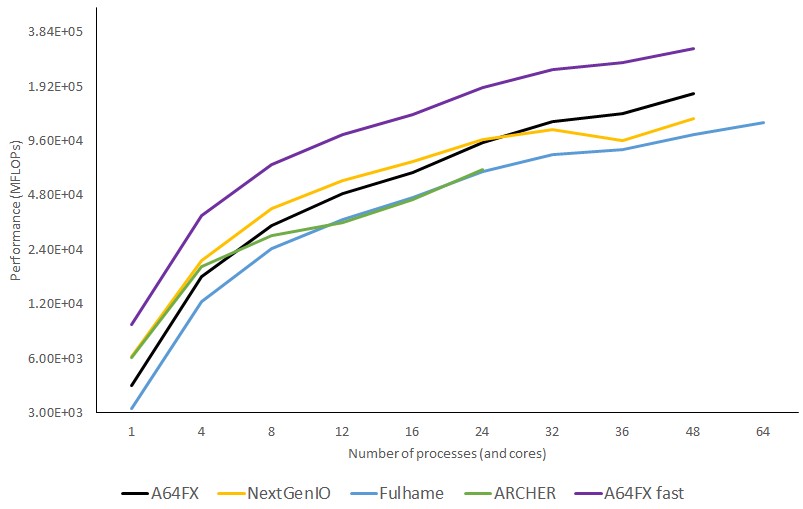}
\caption{Single node scaling across number of cores of a processor (one MPI processes per core)}
\label{fig:nekbone_node_scaling}
\end{figure}

Figure \ref{fig:nekbone_node_scaling} illustrates the performance in MFLOPs on a single node of the different machines across core counts in log scale. It can be seen that the Arm technologies, both the A64FX and ThunderX2 are scaling much better at higher core counts than the Intel technologies, and this in part makes the difference to performance. One can also see that the high core count of the ThunderX2 is a crucial factor here, as at 24 cores it performs comparable to the Ivy Bridge CPUs in ARCHER. 

More generally, it is also interesting that the Ivy Bridge in ARCHER performs very well initially, competitive with the Cascade Lake, but then experiences a significant relative performance decrease beyond four cores.



\subsubsection{Scaling across nodes}

We ran some small inter-node scaling experiments of up to 16 nodes, across the A64FX system, Fulhame and ARCHER. This is interesting for comparison, as Fulhame contains Mellanox EDR IB using a non-blocking fat tree topology, ARCHER is Cray's Dragonfly topology via the Aries interconnect, and the A64FX uses the TofuD network. In all experiments nodes are fully populated, i.e. 48 processes per node on the A64FX, 64 on Fulhame, and 24 on ARCHER. 

Table \ref{fig:nekbonenode-internode-PE} illustrates the inter-node parallel efficiency scaling (defined as the speed up divided by the number of nodes). Nekbone is known to scale well, so it is not surprising that the PEs are so high, although the Infiniband of Fulhame does seem to provide slightly higher performance compared to the other two systems. However this is a simple test, and we have not yet explored the options with the different topologies of the TofuD interconnect (relying simply on the defaults). As such a larger and more challenging test would be instructive, to explore the performance properties of the interconnect more fully.

\begin{table}
\begin{center}
 \caption{Inter-node parallel efficiency across machines}
 \begin{tabular}{|r|r|r|r|} 
 \hline
 Node count & A64FX PE & Fulhame PE & ARCHER PE \\ 
 \hline\hline
 2 & 0.99 & 0.99 & 0.98\\
 4 & 0.97 & 0.99 & 0.98\\
 8 & 0.97 & 0.97 & 0.97\\
 16 & 0.96 & 0.98 & 0.97\\
 \hline
\end{tabular}
\label{fig:nekbonenode-internode-PE}
\end{center}
\end{table}

\section{Applications}

Whilst benchmarks and mini-apps are important for characterising and exploring performance for HPC systems, usability issues and overall system performance evaluation requires running fully functional applications. Therefore, we benchmark using three commonly used HPC applications, two focusing on computational fluid dynamics (COSA and OpenSBLI) and one on materials science (CASTEP). This section outlines those applications and the performance experienced across the systems benchmarked.

\subsection{COSA}
COSA~\cite{COSA_CF} is a CFD application that supports steady, time-domain (TD), and frequency-domain (harmonic balance or HB) solvers, implementing the numerical solution of the NS equations using a finite volume space-discretisation and multigrid (MG) integration.  It is implemented in Fortran and has been parallelised using MPI, with each MPI process working on a set of grid blocks (geometric partitions) of the simulation.  COSA has been shown to exhibit good parallel scaling to large numbers of MPI processes with a sufficiently large test case~\cite{COSA_CSRD}.

\subsubsection{Test Case} \label{sec:cosa_test_case}
The benchmark we used to test the performance of COSA on the A64FX system was a HB test case with 4 harmonics and a grid composed of $800$ blocks, making a total simulation domain of $3,690,218$ grid cells. This was chosen because it fits into approximately 60GB of memory, making it ideal for testing the scaling across a range of nodes on the system. To enable efficient benchmarking the simulation was only run for 100 iterations, a significantly smaller number of iterations than a production run would typically use but enough to evaluate performance sensibly.


\subsubsection{Configuration}

Writing output data to storage can be a significant overhead in COSA, especially for simulations using small numbers of iterations therefore I/O output is disabled to ensure variations in the I/O hardware of the platforms being benchmarked do not affect the performance results collected. The benchmark was run with a single MPI process per core, and all the cores in the node utilised. Table~\ref{tab:cosa-processes-per-node} outlines the cores used per node for the systems benchmarked.

\begin{table}
\begin{center}
 \caption{COSA: Processes per node for each system benchmarked}
 \begin{tabular}{|l|r|r|r|r|r|} 
 \hline
 & A64FX & ARCHER & Cirrus & Fulhame & \makecell[c]{EPCC \\ NGIO} \\ 
 \hline\hline
\makecell[c] {Processes \\ per node}& \makecell[r]{48} & \makecell[r]{24} & \makecell[r]{36} &  \makecell[r]{64} & \makecell[r]{48} \\ 
 \hline
 \end{tabular}
\label{tab:cosa-processes-per-node}
\end{center}
\end{table}

The number of processes used does impact the efficiency of the domain decomposition employed in the application, with best performance exhibited when the number of available decomposition blocks (800 for the test case presented here) exactly divides by the number of processes used. Furthermore, scaling up to all 800 blocks (i.e. using 800 processes) may also introduce some inefficiencies for a small test case such as the one used here, as individual processes may not have enough work to do for optimal performance.

\subsubsection{Results}

\begin{figure}
\centering
\includegraphics[width=\linewidth]{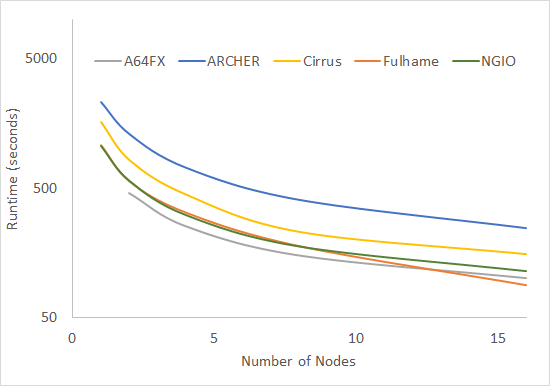}
\caption{COSA performance across a range of nodes counts (strong scaling)}
\label{fig:cosa_nodes_results}
\end{figure}

Figure~\ref{fig:cosa_nodes_results} presents the results of running the benchmark (strong scaling) across a range of node counts for the systems under consideration. Each benchmark was run three times and the average runtime is presented. The benchmark would not fit on a single A64FX node, so the A64FX results start from two nodes. We can see from the graph that the A64FX consistently outperforms the other systems, all the way up to 16 nodes, where performance is overtaken by Fulhame (the ThunderX2 based system). It is worth taking into account the number of MPI processes used on each system, and the number of blocks in the simulation. 

There are 800 blocks in the simulation, meaning at most 800 MPI processes can be active. However, on Fulhame, using 16 nodes the simulation is using 1024 MPI processes, meaning some of the nodes aren't actually undertaking work. For Fulhame only 13 of the nodes are being used, whereas for all the other systems all 16 nodes are active. Furthermore, the number of processes used impacts the load balance, as the data decomposition distributes blocks to processes. Therefore, using 16 nodes on the A64FX will mean there are 800 blocks to be distributed amongst 768 processes, leaving 32 processes with 2 blocks and the rest with 1 block each.  This load imbalance, along with the reduced number of nodes required on Fulhame which minimises the amount of off node MPI communication, is likely to contribute to Fulhame being faster at the highest node count.  

\subsection{CASTEP}
CASTEP \cite{CASTEP, DFT-HK, DFT-KS, RMP-Payne} is a leading code for calculating the properties of materials from first principles. Using density functional theory, it can simulate a wide range of materials proprieties including energetics, structure at the atomic level, vibrational properties, electronic response properties etc. In particular it has a wide range of spectroscopic features that link directly to experiment, such as infra-red and Raman spectroscopies, NMR, and core level spectra.

In this benchmarking we used CASTEP release 18.1.0.  CASTEP requires a high-performance FFT library to function. This is usually provided by FFTW3 or Intel MKL. Fujitsu kindly provided their early development version of FFTW3 for the A64FX platform. CASTEP also requires high-performance BLAS/LAPACK numerical libraries. We used the Fujitsu SSL2 libraries to provide these functions on the A64FX, MKL on the Intel based systems, and the Arm Performance Libraries (Armpl) on the ThunderX2 system.

\subsubsection{Results}

The TiN CASTEP benchmark was run on the different systems with a variety of core counts up to 1 full node and then at various process and thread combinations to use all cores on a node. Note that the benchmark can only be run with total core counts that are either a factor or multiple of 8. This means that on Cirrus, with a core count of 36 cores per node, we cannot use all cores on a node or socket. Instead, we use the number of cores closest to the number of cores available (32 cores for a full node, 16 cores for a socket). For other systems, this means that some combinations of MPI process counts and OpenMP thread counts are impossible. In the majority of combinations we have run the benchmark a minimum of three times and used the best performance from the set of results in each case for comparisons.

\begin{figure}
\centering
\includegraphics[width=\linewidth]{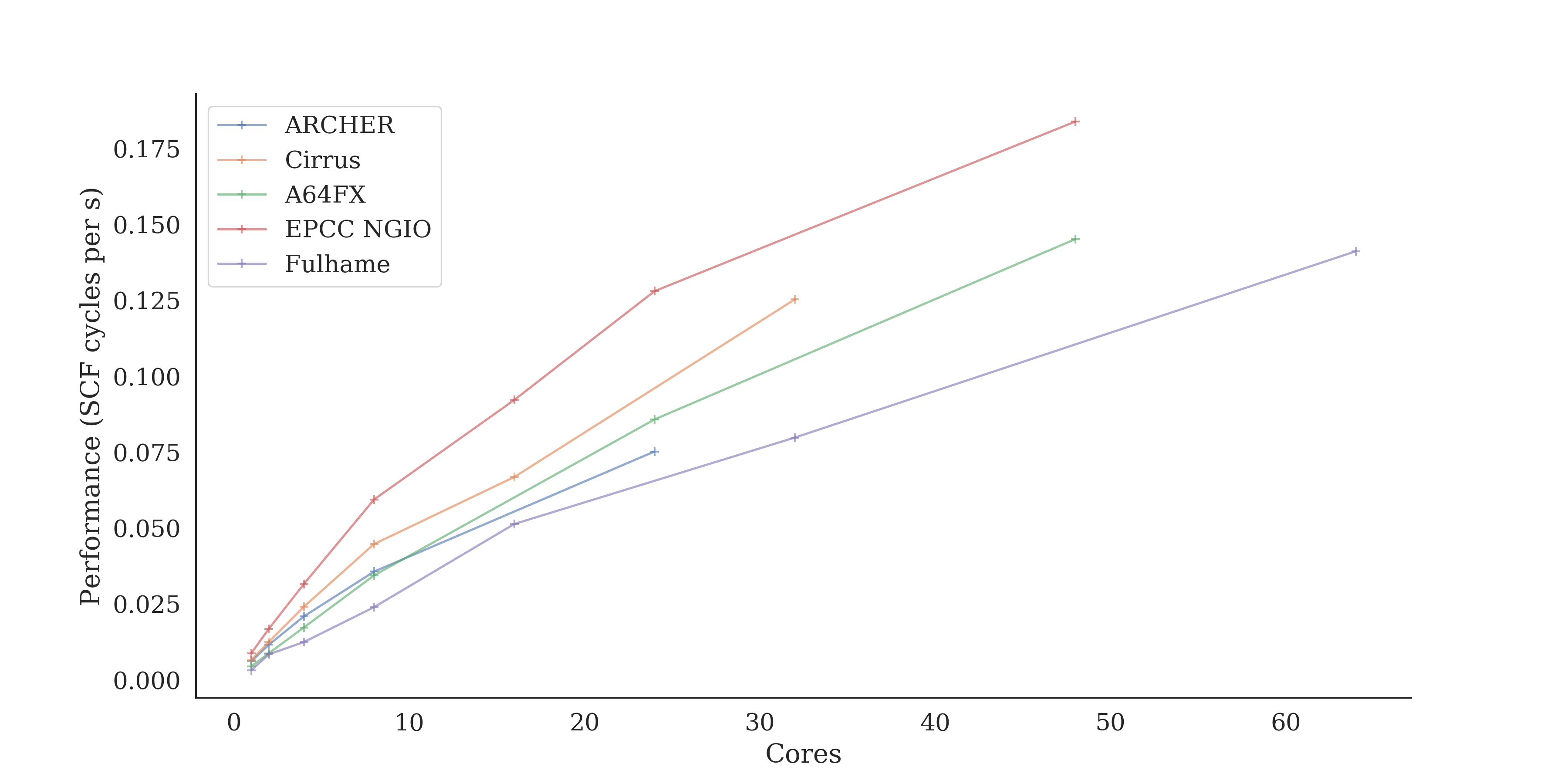}
\caption{Single node CASTEP TiN benchmark performance as a function of core count.}
\label{fig:castep_tin_1node}
\end{figure}

Figure~\ref{fig:castep_tin_1node} shows the performance of CASTEP for the TiN benchmark on 1 node for the test systems. On all systems, the best performance was achieved using MPI only, with no OpenMP threading.

 We can see that the highest absolute single node performance is seen on the EPCC NGIO system with the lowest absolute performance on a single node see on the ARCHER system. Table~\ref{tab:castep-single-node} below shows the performance of the best full-node benchmark runs for each system and the ratio of this performance to the A64FX system.

\begin{table}
 \caption{CASTEP TiN benchmark: best single node performance comparison}
\begin{center}
 \begin{tabular}{|r|r|r|r|} 
 \hline
 System & Cores used & \multicolumn{1}{|c|}{Perf.}  & \multicolumn{1}{|c|}{Ratio to} \\ 
        &            &  \multicolumn{1}{|c|}{(SCF cycles/s)} & \multicolumn{1}{|c|}{A64FX} \\   
 \hline\hline
 A64FX & 48 & 0.145 & 1.00 \\
 ARCHER & 24 & 0.074 & 0.51 \\
 EPCC NGIO & 48 & 0.184 & 1.27 \\
 Cirrus & 32 & 0.125 & 0.86 \\
 Fulhame & 64 & 0.141 & 0.97 \\
 \hline
\end{tabular}
\end{center}
\label{tab:castep-single-node}
\end{table}

The A64FX processor is performing well, providing faster solutions than the ThunderX2 processor, even with lower core counts. However, it is not quite matching the performance of the Intel Cascade Lake processors. As we were working with early versions of FFT libraries, and have yet to attempt A64FX specific optimisations on CASTEP it is likely this performance could be improved, but it is evident that the A64FX processor is competitive in terms of performance for CASTEP.

\subsection{OpenSBLI}

OpenSBLI is a Python-based modelling framework that is capable of expanding a set of differential equations written in Einstein notation, and automatically generating C code that performs the finite difference approximation to obtain a solution. This C code is then targeted with the OPS library towards specific hardware backends, such as MPI/OpenMP for execution on CPUs, and CUDA/OpenCL for execution on GPUs. 

The main focus of OpenSBLI is on the solution of the compressible NS equations with application to shock-boundary layer interactions (SBLI). However, in principle, any set of equations that can be written in Einstein notation can be solve with this framework.

\subsubsection{Test Case}

The benchmark test case setup using OpenSBLI is the Taylor-Green vortex problem in a cubic domain of length $2\pi$. For this study, we have investigated the strong scaling properties for the benchmark on grids of sizes $64\times64\times64$. This is smaller than would normally be run as a benchmark ($512\times512\times512$ and $1024\times1024\times1024$ are common benchmark sizes) but the size is chosen to allow comparisons between single nodes of different architectures as larger benchmarks will not fit into the 32GB available on the A64FX. This benchmark was configured to target pure MPI parallelism and performs minimal I/O.

\subsubsection{Results}

We can see from the results presented in Table~\ref{tab:opensbli_results} that the A64FX  underperforms the other systems, being around 3x and 2x slower than the fastest system (Fulhame). The EPCC NGIO and Fulhame systems present very similar performance, even though they have different characteristics (i.e. EPCC NGIO only has 48 cores and has lower overall memory bandwidth than Fulhame but higher vectorisation capability). For the results presented each test was run three times and the average value is used. There was no significant variations in performance between the individual runs on each system.

\begin{table}[!htbp]
\centering
\caption{OpenSBLI performance (total runtime in seconds)}\label{tab:opensbli_results}
\begin{tabular}{|c|r|r|r|r|}
\hline
\bf{System} & \bf{1 Node} & \bf{2 Nodes} & \bf{4 Nodes} & \bf{8 Nodes} \\
\hline
\hline
A64FX & 3.44 & 1.89 & 1.04 & 0.69 \\
Cirrus & 1.90 & 0.93 & 0.53 & 0.35 \\
EPCC NGIO & 1.18 & 0.75 & 0.46 & 0.31 \\
Fulhame & 1.17 & 0.74 & 0.65 & 0.28 \\
\hline
\end{tabular}
\end{table}

Some initial analysis of OpenSBLI using profiling tools on the A64FX system has shown a large amount of time being spent in both instruction fetch waits and integer cache loads at the L2 cache level. Whilst further investigation is required, and comparative profiling across the range of systems, there is definitely some evidence of potential to optimise this performance with code source code modifications for OpenSBLI.

\section{Conclusions}
We have successfully ported a range of applications and benchmarks to the A64FX processor with minimal effort and no code changes required. This demonstrates the high level of readiness for users of the overall system platform surrounding the A64FX. We have also demonstrated extremely good performance for the A64FX processor for a range of applications, including outperforming other Arm processors and top of the range Intel processors.

However, not all applications exhibit the same performance characterstics on the A64FX processor, with a number of the benchmarks we undertook presenting slightly worse performance, and one benchmark (OpenSBLI) presenting significantly worse performance. When considering this it should be remembered that we have not yet attempted to optimise any of these benchmarks on the target system, aside from using the provided compilers and associated libraries. 
Indeed, the HPCG results demonstrate that on a range of systems there are significant performance benefits that can be achieved by optimising applications for the target architecture. Therefore, we can see that the A64FX processor and computers built with the A64FX technology offer the potential for very significant performance for computational simulation applications.

The benchmarking and evaluation process demonstrated the maturity of the software platform around the A64FX processor (i.e. the compilers, libraries, and batch system). However, it did also highlight that some applications and application domains may struggle with the small amount of memory available on the A64FX-based nodes. This may require work to further parallelise applications, or improve the parallel performance, to exploit additional compute nodes to get sufficient memory for the applications to operate. 
The Fujitsu maths libraries (SSL2) have been shown to be easy replacements for the Intel MKL and Arm performance libraries for some of the applications we have considered in this paper, but not for all requirements we encountered (i.e. FFTW for CASTEP). Therefore, some further work on optimised libraries for the A64FX system would be beneficial.


\section*{Acknowledgments}
Access to the A64FX was provided through the Fujitsu early access programme. The EPCC NGIO system was funded by the European Union's Horizon 2020 Research and Innovation programme under Grant Agreement no. 671951. The Fulhame HPE Apollo 70 system is supplied to EPCC as part of the Catalyst UK programme, a collaboration with Hewlett Packard Enterprise, Arm and SUSE to accelerate the adoption of Arm based supercomputer applications in the UK. This work used the Cirrus UK National Tier-2 HPC Service at EPCC (http://www.cirrus.ac.uk) funded by the University of Edinburgh and EPSRC (EP/P020267/1). This work used the ARCHER UK National Supercomputing Service (http://www.archer.ac.uk).



%


\IEEEtriggeratref{21}

\bibliographystyle{IEEEtran}
\bibliography{eahpc}

\begin{thebibliography}{10}
\providecommand{\url}[1]{#1}
\csname url@samestyle\endcsname
\providecommand{\newblock}{\relax}
\providecommand{\bibinfo}[2]{#2}
\providecommand{\BIBentrySTDinterwordspacing}{\spaceskip=0pt\relax}
\providecommand{\BIBentryALTinterwordstretchfactor}{4}
\providecommand{\BIBentryALTinterwordspacing}{\spaceskip=\fontdimen2\font plus
\BIBentryALTinterwordstretchfactor\fontdimen3\font minus
  \fontdimen4\font\relax}
\providecommand{\BIBforeignlanguage}[2]{{%
\expandafter\ifx\csname l@#1\endcsname\relax
\typeout{** WARNING: IEEEtran.bst: No hyphenation pattern has been}%
\typeout{** loaded for the language `#1'. Using the pattern for}%
\typeout{** the default language instead.}%
\else
\language=\csname l@#1\endcsname
\fi
#2}}
\providecommand{\BIBdecl}{\relax}
\BIBdecl

\bibitem{top500_fugaku}
\BIBentryALTinterwordspacing
Top500 fugaku number one. [Online]. Available:
  \url{https://www.top500.org/news/japan-captures-top500-crown-arm-powered-supercomputer/}
\BIBentrySTDinterwordspacing

\bibitem{green500_fugaku}
\BIBentryALTinterwordspacing
Green500 june 2020. [Online]. Available:
  \url{https://www.top500.org/lists/green500/2020/06/}
\BIBentrySTDinterwordspacing

\bibitem{tofud_network}
Y.~{Ajima}, T.~{Kawashima}, T.~{Okamoto}, N.~{Shida}, K.~{Hirai}, T.~{Shimizu},
  S.~{Hiramoto}, Y.~{Ikeda}, T.~{Yoshikawa}, K.~{Uchida}, and T.~{Inoue}, ``The
  tofu interconnect d,'' in \emph{2018 IEEE International Conference on Cluster
  Computing (CLUSTER)}, 2018, pp. 646--654.

\bibitem{Arm-PASC19}
\BIBentryALTinterwordspacing
A.~Jackson, A.~Turner, M.~Weiland, N.~Johnson, O.~Perks, and M.~Parsons,
  ``Evaluating the arm ecosystem for high performance computing,'' in
  \emph{Proceedings of the Platform for Advanced Scientific Computing
  Conference}, ser. PASC ’19.\hskip 1em plus 0.5em minus 0.4em\relax New
  York, NY, USA: Association for Computing Machinery, 2019. [Online].
  Available: \url{https://doi.org/10.1145/3324989.3325722}
\BIBentrySTDinterwordspacing

\bibitem{10.1007/978-3-030-43229-4_17}
E.~Calore, A.~Gabbana, F.~Rinaldi, S.~F. Schifano, and R.~Tripiccione, ``Early
  performance assessment of the thunderx2 processor for lattice based
  simulations,'' in \emph{Parallel Processing and Applied Mathematics},
  R.~Wyrzykowski, E.~Deelman, J.~Dongarra, and K.~Karczewski, Eds.\hskip 1em
  plus 0.5em minus 0.4em\relax Cham: Springer International Publishing, 2020,
  pp. 187--198.

\bibitem{SPECMPI}
M.~Müller, M.~van Waveren, R.~Lieberman, B.~Whitney, H.~Saito, K.~Kumaran,
  J.~Baron, W.~Brantley, C.~Parrott, T.~Elken, H.~Feng, and C.~Ponder, ``Spec
  mpi2007-an application benchmark suite for parallel systems using mpi,''
  \emph{Concurrency and Computation: Practice and Experience}, vol.~22, pp.
  191--205, 02 2010.

\bibitem{HPCChallenge}
J.~Dongarra and P.~Luszczek, ``Introduction to the hpcchallenge benchmark
  suite,'' 12 2004.

\bibitem{NASBenchmark}
D.~Bailey, J.~Barton, T.~Lasinski, and H.~Simon, ``The nas parallel
  benchmarks,'' 08 1993.

\bibitem{ARCHER_Benchmarks}
\BIBentryALTinterwordspacing
Archer benchmarks. [Online]. Available:
  \url{http://www.github.com/hpc-uk/archer-benchmarks}
\BIBentrySTDinterwordspacing

\bibitem{ARCHER_bench_perf_report}
\BIBentryALTinterwordspacing
A.~Turner and J.~Salmond, ``{hpc-uk/archer-benchmarks: Initial performance
  comparison report},'' Jun. 2018. [Online]. Available:
  \url{https://doi.org/10.5281/zenodo.1288378}
\BIBentrySTDinterwordspacing

\bibitem{A64FX_hot_chips}
T.~Yoshida, ``Fujitsu high performance cpu for the post-k computer,'' in
  \emph{Hot Chips 30 Symposium (HCS)}, ser. Hot Chips, 2018.

\bibitem{A64FX_sim}
\BIBentryALTinterwordspacing
Y.~Kodama, T.~Odajima, A.~Asato, and M.~Sato, ``Evaluation of the riken post-k
  processor simulator,'' 04 2019. [Online]. Available:
  \url{https://arxiv.org/abs/1904.06451}
\BIBentrySTDinterwordspacing

\bibitem{Dongarra15HPCG}
J.~J. Dongarra, M.~A. Heroux, and P.~Luszczek, ``Hpcg benchmark : a new metric
  for ranking high performance computing systems,'' Knoxville, Tennessee, Tech.
  Rep. UT-EECS-15-736, November 2015.

\bibitem{HPL}
J.~J. Dongarra, P.~Luszczek, and A.~Petitet, ``The linpack benchmark: past,
  present and future,'' \emph{Concurrency and Computation: Practice and
  Experience}, vol.~15, pp. 803--820, 2003.

\bibitem{karp2020optimization}
M.~Karp, N.~Jansson, A.~Podobas, P.~Schlatter, and S.~Markidis, ``Optimization
  of tensor-product operations in nekbone on gpus,'' \emph{arXiv preprint
  arXiv:2005.13425}, 2020.

\bibitem{COSA_CF}
W.~Jackson, M.~Campobasso, and J.~Drofelnik,
  ``\BIBforeignlanguage{English}{Load balance and parallel i/o: Optimising cosa
  for large simulations},'' \emph{\BIBforeignlanguage{English}{Computers and
  Fluids}}, 3 2018.

\bibitem{COSA_CSRD}
A.~Jackson and M.~Campobasso, ``\BIBforeignlanguage{English}{Shared-memory,
  distributed-memory, and mixed-mode parallelisation of a cfd simulation
  code},'' \emph{\BIBforeignlanguage{English}{Computer Science - Research and
  Development}}, vol.~26, no. 3-4, pp. 187--195, 6 2011.

\bibitem{CASTEP}
S.~J. Clark, M.~D. Segall, C.~J. Pickard, P.~J. Hasnip, M.~J. Probert,
  K.~Refson, and M.~Payne, ``First principles methods using {CASTEP},''
  \emph{Z. Kristall.}, vol. 220, pp. 567--570, 2005.

\bibitem{DFT-HK}
P.~Hohenberg and W.~Kohn, ``Inhomogeneous electron gas,'' \emph{Phys. Rev.},
  vol. 136, pp. B864--B871, 1964.

\bibitem{DFT-KS}
W.~Kohn and L.~J. Sham, ``Self-consistent equations including exchange and
  correlation effects,'' \emph{Phys. Rev.}, vol. 140, pp. A1133--A1138, 1965.

\bibitem{RMP-Payne}
M.~C. Payne, M.~P. Teter, D.~C. Allan, T.~Arias, and J.~D. Joannopoulos,
  ``Iterative minimization techniques for ab initio total-energy calculations -
  molecular-dynamics and conjugate gradients,'' \emph{Rev. Mod. Phys.},
  vol.~64, pp. 1045--1097, 1992.

\end{thebibliography}

\end{document}